\documentstyle{tmu}
\input{psfig}
\def\epsfile#1{\leavevmode\psfig{#1}}
\makeatletter
\ifx\undefined\@p@@sscale \def\@p@@sscale {\errmessage{scale=  unspported}}\fi
\ifx\undefined\@p@@svscale\def\@p@@svscale{\errmessage{vscale= unspported}}\fi
\ifx\undefined\@p@@shscale\def\@p@@shscale{\errmessage{hscale= unspported}}\fi
\makeatother

\def\be{\begin{equation}}
\def\ee{\end{equation}}
\def\bea{\begin{eqnarray}}
\def\eea{\end{eqnarray}}
\def\simm#1{\mathop{\vtop{\ialign{##\crcr
        $\hfil\displaystyle{#1}\hfil$\crcr\noalign{\kern0.5pt\nointerlineskip}
        $\sim$\crcr\noalign{\kern0.5pt}}}}\limits}

\begin{document}

\title{%
\begin{flushright}
\vspace*{-8mm}
{\normalsize
UTCCP-P-84\\
UTHEP-421\\
April 2000\\
}
\end{flushright}
Large-Flavor QCD on the Lattice\footnote{%
Presented by K. Kanaya
at TMU-Yale Symposium on 
{\it Dynamics of Gauge Fields} --- an External Activity of APCTP,
Dec.\ 13--15, 1999, Hachioji, Japan
}
}

\author{%
      Y. Iwasaki, K. Kanaya, S. Kaya, T. Yoshi\'e\\
{\it  Institute of Physics (Center for Computational Physics),
University of Tsukuba, Tsukuba, Ibaraki 305-8577, Japan.
iwasaki@rccp.tsukuba.ac.jp,\\
kanaya@rccp.tsukuba.ac.jp,
kaya@imslab.co.jp,
yoshie@rccp.tsukuba.ac.jp
}\\
      S. Sakai\\
{\it  Faculty of Education, Yamagata University, Yamagata 990-8560, Japan.\\
sakai@kecszao.kj.yamagata-u.ac.jp
}
}

\maketitle

\section*{Abstract}

We study the nature of the QCD vacuum at 
general number of flavors $N_F$ by numerical simulations on the lattice.
Combining the results with those of the perturbation theory, 
we propose the following picture:
1) For $N_F \ge 17$, there exists only
one IR fixed point at vanishing gauge coupling, i.e.,
the theory in the continuum limit is trivial.
2) For $16 \ge N_F \ge 7$, there is a non-trivial fixed point.
Therefore, the theory is non-trivial with anomalous dimensions,
however, without quark confinement.
3) For $N_F \le 6$, theories satisfy both quark confinement and 
spontaneous chiral symmetry breaking in the continuum limit.

\section{Introduction}
\label{sect:intro}

The beta function of QCD is universal up to two-loop in the 
perturbation theory:
$
\tilde{\beta}(g) = - b_0 g^3 - b_1 g^5,
$
with
$
b_0 = \frac{1}{16\pi^2} \left( 11-\frac{2}{3}N_F \right)
$
and
$
b_1 = \frac{1}{(16\pi^2)^2} \left( 102-\frac{38}{3}N_F \right).
$
When the number of flavors ($N_F$) exceeds 
$16 \frac{1}{2}$, $b_0$ becomes negative, so that 
the asymptotic freedom of QCD is lost.
In this case, the origin $g=0$ becomes an IR fixed point, 
i.e. the low energy limit of QCD is a free theory. 
Quarks cannot be confined and the chiral symmetry cannot be broken
spontaneously.
$N_F > 16 \frac{1}{2}$ is a sufficient, but not a necessary condition for these.
A natural question is, whether confinement and spontaneous breaking of 
the chiral symmetry are satisfied at all $N_F$ below $16 \frac{1}{2}$.

It is well-known that the second coefficient $b_1$ changes its sign 
already at $N_F \approx 8.05$. 
Therefore, a non-trivial IR fixed point may appear at finite $g$.
From the two-loop beta function, this happens for $N_F = 9$--16.
At least for $N_F \sim 16$, the IR fixed point shows up in a 
perturbative region. Therefore,
it is plausible that the full beta function has a non-trivial 
IR fixed point for $N' \leq N_F \leq 16$ with some $N' \le 16$.
When such an IR fixed point exists, the coupling constant cannot become 
arbitrarily large in the IR region --- 
this will imply that quarks are not confined.
In particular, for $N_F \sim 16$, QCD is perturbative also in the 
IR limit, i.e., we cannot expect confinement.

In this paper, we study the $N_F$-dependence of the QCD vacuum. 
A non-perturbative investigation is required.
Lattice QCD is the only systematic method that enables us to
study non-perturbative properties of QCD.
We performed a series of lattice simulations for a wide range of $N_F$,
from 2 to 300,
to study the phase structure of QCD.
When the phase diagram becomes clear, 
we are able to see the nature of the QCD vacuum in the continuum limit,
and eventually answer the question about the condition on $N_F$ 
for confinement and spontaneous breakdown of the chiral symmetry.

\begin{figure}[tb]
\begin{center}
a)\epsfile{file=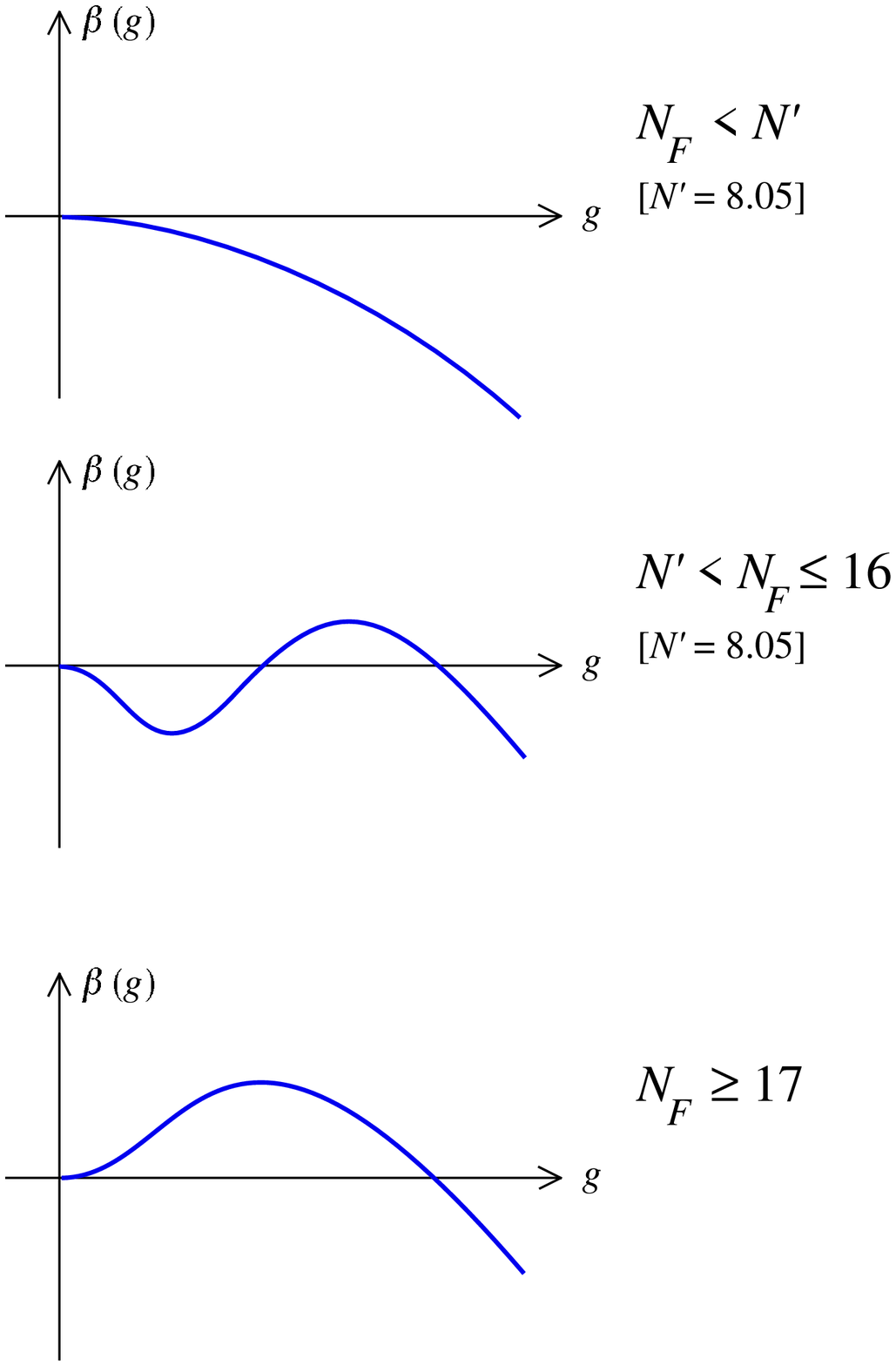,width=0.4\textwidth}
\makebox[5mm]{}
b)\epsfile{file=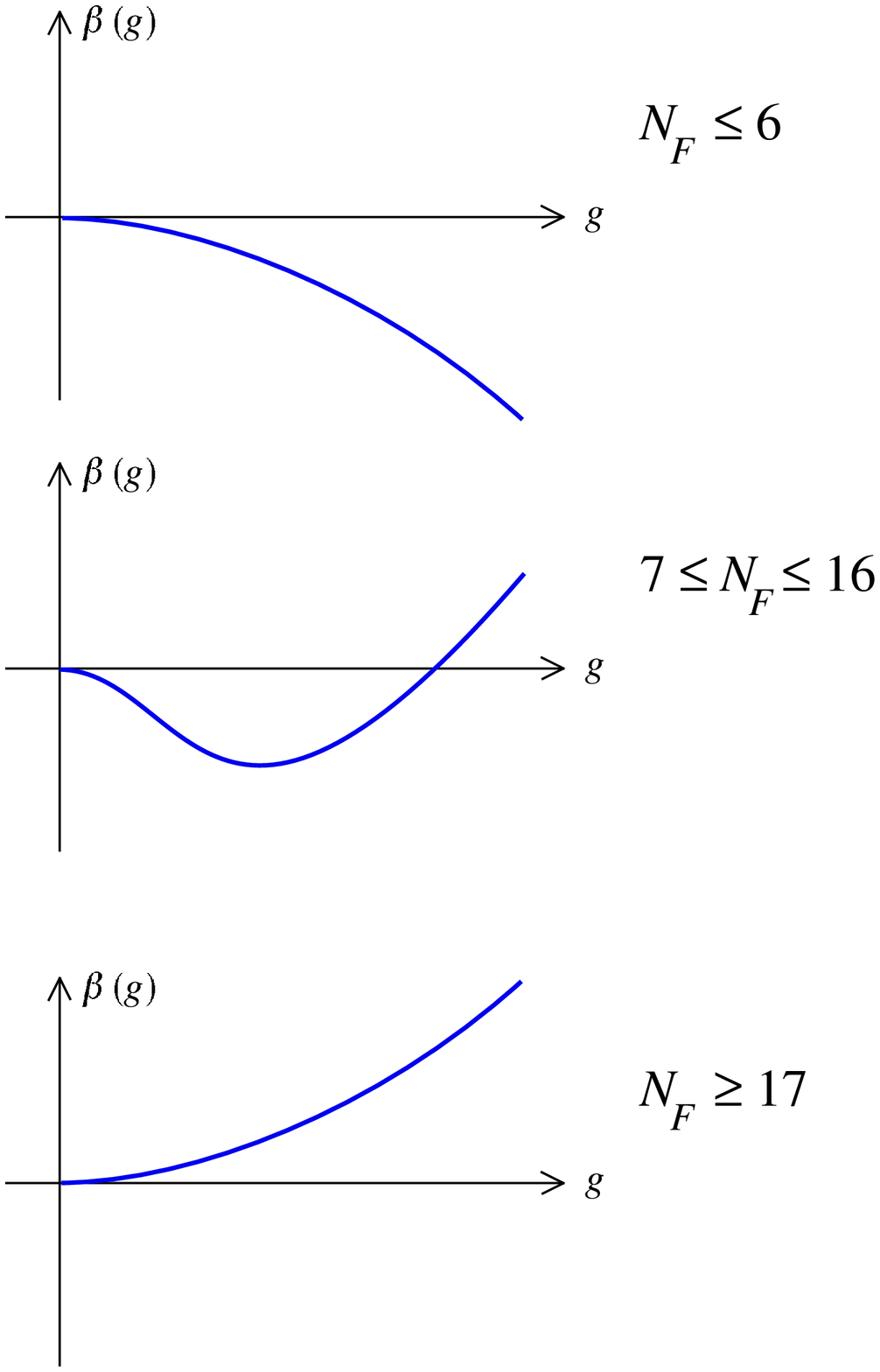,width=0.4\textwidth}
\end{center}
\vspace{-0.5cm}
\caption{Renormalization group beta function of QCD.
(a) Conjecture by Banks and Zaks \protect\cite{Banks1} 
assuming confinement in the strong coupling limit for all $N_F$.
(b) Our conjecture deduced from the results of lattice simulations.
}
\label{BetaFunc}
\end{figure}

In 1982, Banks and Zaks published a pioneering paper on the $N_F$-dependence 
of the QCD vacuum \cite{Banks1}.
Based on an early result obtained on the lattice \cite{Kogut79},
they assumed that, in the strong coupling limit, the theory is confining 
and the beta function is negative, for any value of $N_F$. 
Using the perturbative results for the beta function discussed above, 
they conjectured Fig.~\ref{BetaFunc}(a) as the simplest $N_F$ dependence
of the beta function, 
and studied the phase structure of QCD based on this beta function. 
The assumption of negative beta function in the strong coupling region
leads to an additional non-trivial UV fixed point for $N_F \ge N'$. 
Due to this UV fixed point, their conjecture for the phase structure is 
complicated.
[For different approaches see Refs.~\cite{Oehme,Nishijima,Appelquist,sigma}.]

In the arguments of Banks and Zaks,
the assumption of confinement and negative beta function in the strong 
coupling limit plays an essential role.
Here, it should be noted that the lattice study on which this assumption
was based, is a study of a pure gauge theory \cite{Kogut79}.
With dynamical quarks, the vacuum structure can be much more complicated. 
Actually, in the realistic cases, 
there exist no proofs of confinement for general $N_F$ even 
in the strong coupling limit.%
\footnote{%
We can formally rewrite the theory in terms of mesons and baryons.
We may argue that, when the resulting effective action for hadrons 
is a well-defined action of weakly interacting particles,
it is plausible that quarks are confined --- the hadrons are 
good variables to describe the dynamical contents of the theory.
In the strong coupling limit,
the effective action can be computed either in the large $N_c$ 
limit \cite{SCENc},
using a meanfield approximation ($1/d$ expansion) \cite{SCEmf,SCEmfMq},
or using a heavy quark mass expansion \cite{SCEmfMq,wilson77}.
In these cases, the effective action for mesons and baryons seems to lead 
to a well-defined world with spontaneously broken chiral symmetry,
suggesting quark confinement.
For the realistic cases of $N_c=3$ with light quarks, however, 
the effective action is quite complicated;
diagrams with quark loops, which are suppressed at large $N_c$,
large $d$ or large $m_q$, become important.
Therefore, in the realistic cases, 
it is difficult to analytically deduce the vacuum structure 
as a function of $N_F$ even in the strong coupling limit.
}
A non-perturbative investigation on the lattice is required.

In a previous study, we performed simulations of QCD 
in the strong coupling limit at various $N_F$ \cite{previo}. 
We found that, when $N_F \ge 7$, 
light quarks are {\it not} confined and chiral
symmetry is restored, even in the strong coupling limit.
Here, we extend this study to weaker couplings and to a wider range of $N_F$
up to 300. 
Combining the lattice results and perturbative arguments,
we conjecture Fig.~\ref{BetaFunc}(b) for the non-perturbative beta 
function of QCD.
This leads to the following $N_F$-dependence of the QCD vacuum.
\begin{itemize}
\item
When $N_F$ is smaller than a critical value, 
the beta function is negative for all values of $g$.
Quarks are confined and the chiral symmetry is spontaneously broken
at zero temperature.
We find that the critical number of flavors is 6.
(Corresponding critical value of $N_F$ is 2 for $N_c=2$, 
to be compared with the two-loop value 5 \cite{lat9496}.)
\item
On the other hand, when $N_F$ is equal or larger than 17,
we conjecture that the beta function is positive for all $g$,
in contrast to 
the conjecture by Banks and Zaks shown in Fig.~\ref{BetaFunc}(a).
The theory is trivial in this case.
\item
When $N_F$ is between 7 and 16, the beta function changes 
sign from negative to positive with increasing $g$. 
Therefore, the theory has a non-trivial IR fixed point,
i.e., the theory is non-trivial with anomalous 
dimensions.  We conjecture that quarks are not confined in this case.
\end{itemize}
Our results are in part reported in Refs.~\cite{lat9496} and \cite{YKIS97}.

This report is organized as follows:
In Sec.~\ref{sec:model}, 
we describe our lattice model and simulation parameters.
Results for the phase structure in the strong coupling limit are given 
in Sec.~\ref{sec:strong}.
The phase structure at finite coupling is discussed in 
Sec.~\ref{sec:finitebeta}.
%We then study the nature of the deconfined phase in 
%Sec.~\ref{sec:deconf}.
Finally, in Sec.~\ref{sec:conclusions}, we summarize our conclusions 
and conjectures.

%%%%%%%%%%%%%%%%%%%%%%%%%%%%%%%%%%%%%%%%%%%%%%%%%%%%%%%%%%%%%%%%%%%%%%
\section{Model and simulation parameters}
\label{sec:model}

We consider a 4-dimensional Euclidian lattice with the lattice 
spacing $a$ and finite lattice volume $(N_s a)^3 \times (N_t a)$,
where $N_s$ and $N_t$ are the number of lattice sites in spatial and temporal
directions. The action consists of the gauge part and the quark part,
$ S = S_{gauge} + S_{quark}$.
For the gauge part, we use the standard one plaquette action.
\be
S_{gauge} = - \frac{\beta}{6} \sum_{x,\mu\neq\nu}{\rm Tr}
\left( U_{x,\mu}U_{x+\hat\mu,\nu}
U_{x+\hat\mu+\hat\nu,\mu}^{\dag} U_{x+\hat\nu,\nu}^{\dag}\right),
\label{eqn:gaction}
\ee
where $U_{x,\mu}$ is the gauge connection between the site $x$ and the 
neighboring site in the positive $\mu$ direction, and $\beta=6/g^2$.

For the quark part, there are several alternatives. 
Two conventional choices are the staggered (Kogut-Susskind) 
fermion action \cite{susskind77} 
and the Wilson fermion action \cite{wilson77}.  
Staggered quark has been preferably used in previous studies\cite{KS} 
because a part of the chiral symmetry is preserved in this formulation.
On the other hand, the flavor structure is quite complicated:
The action is local only when $N_F$ is a multiple of 4. 
Off the continuum limit, there exists a lattice artifact mixing 
these 4 degenerate flavors.
For the more realistic cases $N_F=2$ and 3, 
we modify by hand the power of the fermionic determinant in the 
numerical path-integration. 
This necessarily makes the action non-local, that sometimes 
poses conceptually and technically difficult problems.

Another conventional choice for the quark action is the Wilson 
fermion action, which we adopt in the followings.
\be
S_{quark} = \sum_{f,x} \left[ \bar{\psi}_x^f \psi_x^f
-K\left\{\bar{\psi}_{x}^f(1\!-\!\gamma_{\mu})U_{x,\mu}{\psi_{x+\hat{\mu}}^f}
+\bar{\psi}_{x+\hat{\mu}}^f(1\!+\!\gamma_{\mu})U_{x,\mu}^{\dag}\psi_{x}^f
 \right\} \right],
\label{eqn:wilson}
\ee
The index $f$ ($=1, \cdots, N_F$) is for the flavors.
We assume that all $N_F$ quarks are degenerate with the bare mass $m_0$, 
which is related to the hopping parameter $K$ by 
$ m_0 = (1/2a) (1/K-1/K_c) $, 
where $K_c=1/8$ is the point where the bare mass vanishes.
%The point $K=0$ corresponds to infinite quark mass.
In this formulation, the chiral symmetry is explicitly broken at finite $a$. 
On the other hand, 
the flavor symmetry as well as the C, P and T symmetries are manifestly
satisfied without mixing etc.\ also on the lattice. 

In this study, we use the Wilson fermion formalism
because this is the only formalism that preserves manifest flavor symmetry. 
This feature will be important in a study of $N_F$-dependence in QCD.
Concerning the chiral symmetry, 
a perturbative study of Ward identities shows 
that the effects of the $O(a)$ chiral breaking terms 
can be absorbed by appropriate renormalizations, 
including an additive renormalization of quark mass \cite{Bo}.
As a result, the location of the chiral limit $K_c$ shifts from the
tree value 1/8 as a function of $\beta$.

In order to take the additive mass renormalization into account,
we define the current quark mass $m_q$ in terms of an axial vector
Ward identity \cite{ItohNP,Bo}. 
\be
2 m_q \langle\,0\,|\,P\,|\,\pi(\vec{p}=0)\,\rangle
= - m_\pi \, \langle\,0\,|\,A_4\,|\,\pi(\vec{p}=0)\,\rangle
\label{eq:mq}
\ee
where $P$ is the pseudoscalar density, $A_4$ the fourth component of the
local axial vector current, and $m_\pi$ the pion (screening) mass. 
A multiplicative renormalization factor for the axial current,
which is not important in this study, is absorbed into the definition 
of the quark mass.
We then define $K_c$ by the condition $m_q=0$.
Numerical simulations for small $N_F$ shows that $K_c(\beta)$ is a smooth 
curve connecting 1/8 at $\beta = \infty$ and 1/4 at $\beta =0$.

In QCD, even when the confinement is realized at low temperatures, 
we expect deconfinement at high temperatures \cite{KanayaYKIS97}.
Therefore, we have to study the temperature dependence of
the results to distinguish between the finite temperature 
deconfinement transition/crossover due to high temperatures 
and a bulk transition at zero temperature due to the effects of many flavors.
On a lattice with $N_t$ sites in the Euclidian time direction,
the temperature is given by $T=1/N_t a$.
When the beta function is negative, as in the case of small $N_F$,
the lattice spacing $a$ is a decreasing function of $\beta=6/g^2$.
In this case, $T$ increases with increasing $\beta$ 
for a fixed finite $N_t$.
On the other hand, when the beta function is positive, 
$T$ decreases with increasing $\beta$.
Therefore, in order to study the temperature dependence for general $N_F$,
we have to compare the results at various $N_t$.

Numerical study shows that the value of $m_q$ defined through the axial vector 
Ward identity does not depend on
whether the system is in the high or the low temperature phase
when $\beta$ is large; $\beta \simm{>} 5.5$ for $N_F=2$ \cite{Tsukuba91}.
When we define the pion decay constant $f_\pi$ by 
$ %%\be
\langle\,0\,|\,A_4\,|\,\pi(\vec{p}=0)\,\rangle = m_\pi f_\pi,
%%\label{eq:fpi}
$ %%\ee
either $m_\pi=0$ or $f_\pi=0$ has to be satisfied in the chiral limit.
The chiral symmetry is restored when $f_\pi=0$ in the chiral limit,
while $m_\pi=0$ if the chiral symmetry is spontaneously broken \cite{Stand26}.

We perform simulations on lattices $8^2 \times 10 \times N_t$ 
($N_t =4$, 6, and 8), $16^2 \times 24 \times N_t$ ($N_t=16$)
and $18^2 \times 24 \times N_t$ ($N_t=18$).
We vary $N_F$ from 2 to 300.
For each $N_F$, we study the phase structure in the coupling 
parameter space $(\beta,K)$.
It should be noted that, in QCD with dynamical quarks, 
there are no order parameters for quark confinement. 
We discuss about confinement by comparing the screening pion mass 
and the lowest Matsubara frequency, and, simultaneously, consulting
the values of plaquette and the Polyakov loop.
In the followings, we call the pion screening mass simply 
the pion mass, and similarly for the quark mass.
Further details about our simulation are described in 
Refs.~\cite{lat9496,YKIS97}

%%%%%%%%%%%%%%%%%%%%%%%%%%%%%%%%%%%%%%%%%%%%%%%%%%%%%%%%%%%%%%%%%%%
\section{Strong coupling limit $\beta=0$}
\label{sec:strong}

\begin{figure}[tb]
\begin{center}
a)\epsfile{file=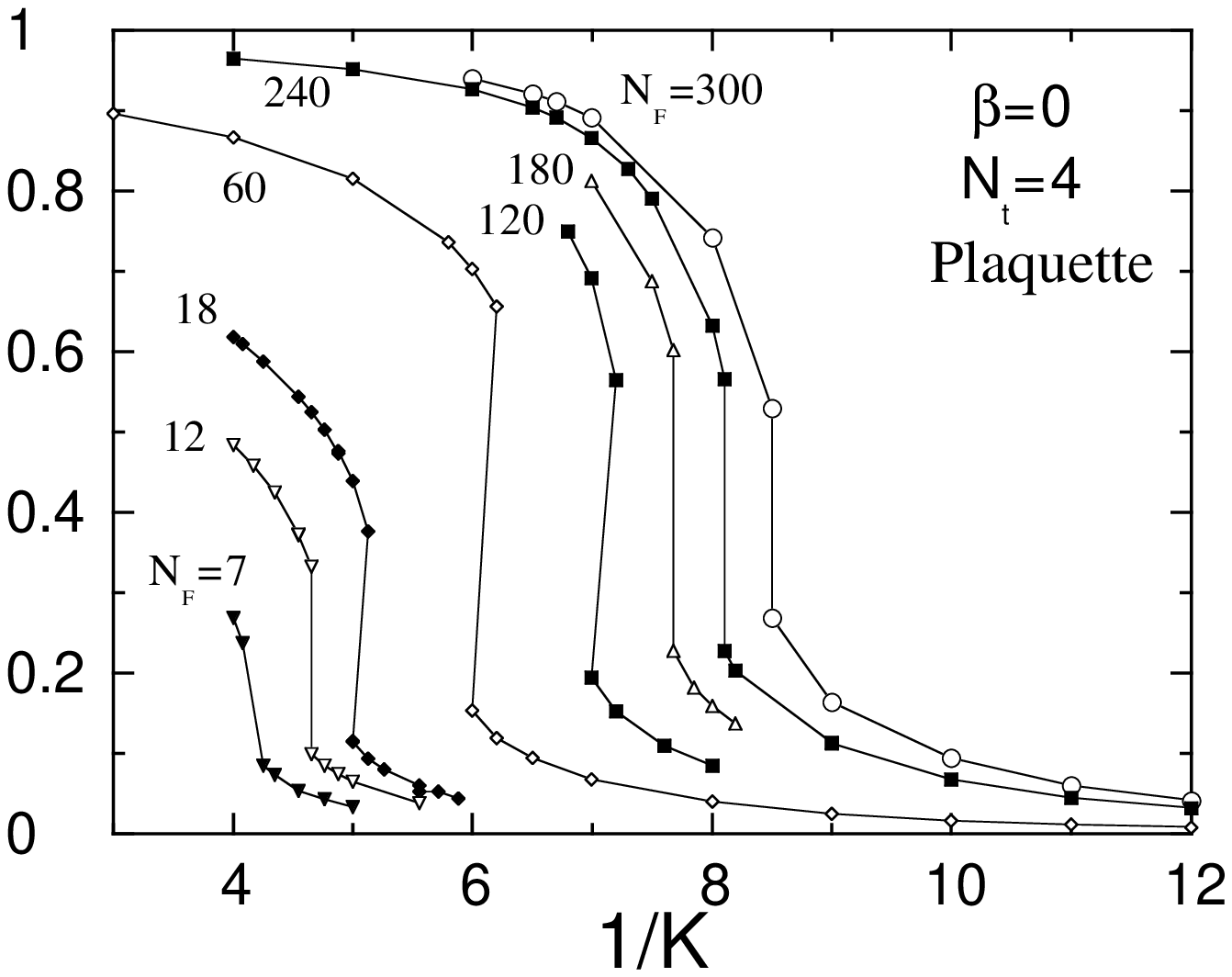,width=0.45\textwidth}
%\makebox[5mm]{}
b)\epsfile{file=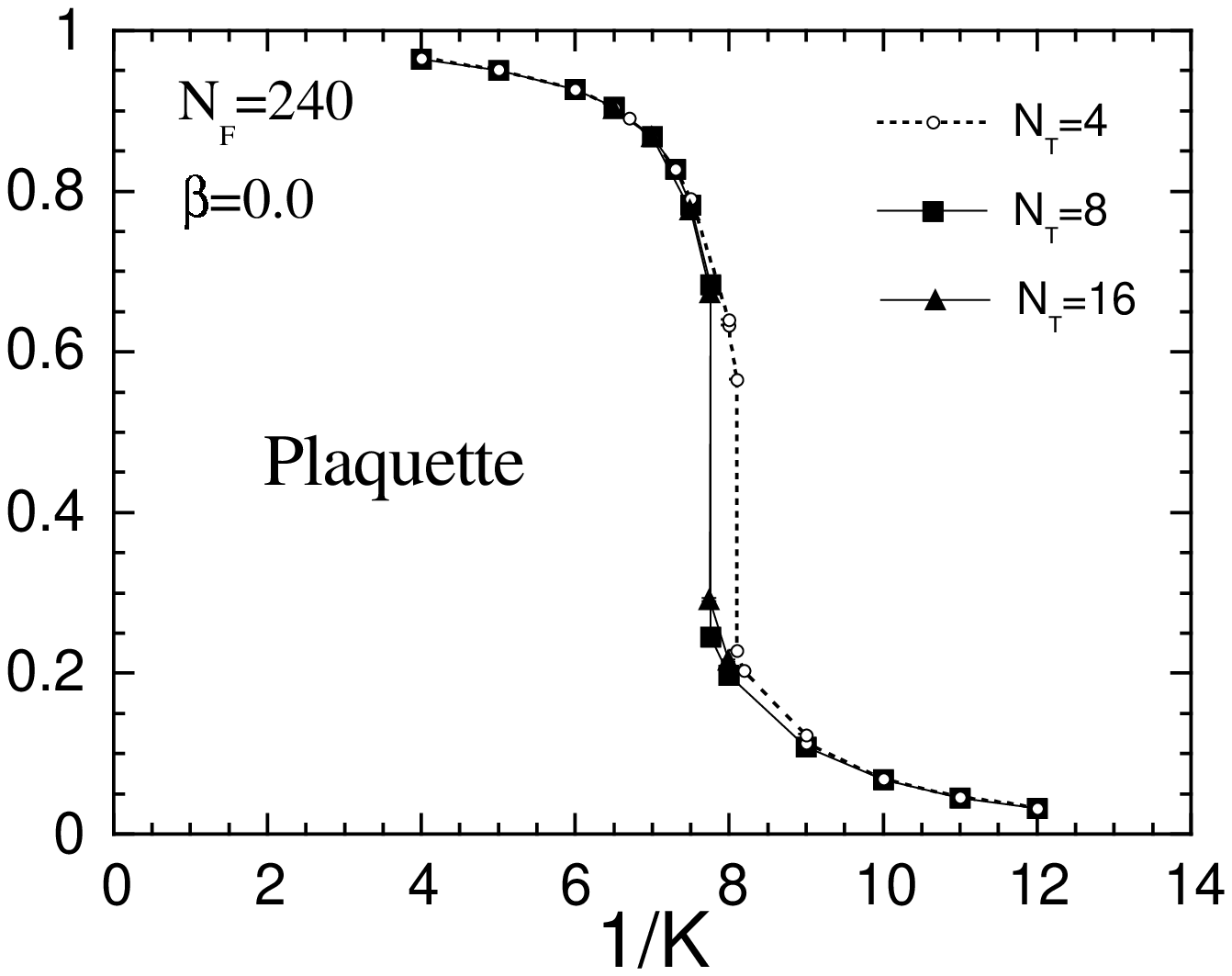,width=0.45\textwidth}
\end{center}
\vspace{-0.5cm}
\caption{
Plaquette at $\beta=0$ as a function of $1/K$.
(a) For $N_F=7$--300 at $N_t=4$.
(b) For $N_F=240$ at $N_t=4$--16.
}
\label{B0.W}
\end{figure}

In a previous paper \cite{previo}, we have shown that, when $N_F \ge 7$, 
light quarks are deconfined and chiral symmetry is restored 
at zero temperature even in the strong coupling limit.
For $N_F \leq 6$, we have only one confining phase from the heavy
quark limit $K=0$ up to the chiral limit $K_c=0.25$.
On the other hand, 
for $N_F \geq 7$, we find a strong first order transition at 
$K=K_d$:
\begin{itemize}
\item
When quarks are heavy ($K < K_d$), both plaquette and the Polyakov 
loop are small, and $m_\pi$ satisfies the PCAC relation 
$m_\pi^2 \propto m_q$. 
Therefore, quarks are confined and the chiral 
symmetry is spontaneously broken in this phase.
We find that $m_q$ in the confining phase is non-zero at the 
transition point $K_d$. 
While an extrapolation in $1/K$ shows that $m_{\pi}^2$ decreases 
towards zero at $K=0.25$, the chiral limit does not belong to this phase
($K_d < K_c$).
\item
On the other hand, when quarks are light ($K > K_d$), plaquette and 
the Polyakov loop are large.
In this phase, $m_\pi$ remains large in the chiral limit
and is almost equal to twice the lowest Matsubara frequency $\pi/N_t$. 
This implies that the pion state is a free two-quark state 
and, therefore, quarks are not confined in this phase.
The pion mass is nearly equal to the scalar meson mass, 
and the rho meson mass to the axial vector meson mass. 
The chiral symmetry is also manifest within
corrections due to finite lattice spacing.
\end{itemize}
See \cite{previo} for details.

\begin{figure}[tb]
\begin{center}
\epsfile{file=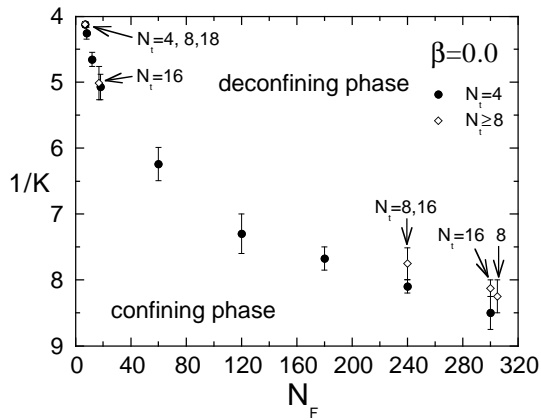,width=0.5\textwidth}
\end{center}
\vspace{-0.5cm}
\caption{
The transition point $1/K_d$ at $\beta=0$ versus $N_F$ 
for $N_t=4$ and $N_t\ge 8$. 
For clarity, data at $N_t=8$ for $N_F=300$ is slightly shifted 
to a larger $N_F$ in the figure. 
}
\label{Kd.Nf}
\end{figure}

We extend the study to larger $N_F$. 
We show the results of plaquette 
at $N_t=4$ for $N_F =7$--300 in Fig.~\ref{B0.W}(a).
Clear first order transition can be seen at $K$ below 0.25 ($1/K > 4$).
We then study the $N_t$ dependence of the results, 
as shown in Fig.~\ref{B0.W}(b) for the case of $N_F=240$.
We find that,
although the transition point shows a slight shift to smaller $1/K$
when we decrease $T=1/N_t a$ from $N_t=4$ to 8, it stays at 
the same point for $N_t \geq 8$.
%($1/K_d \simeq 8.1(1)$ for $N_t=4$ and $1/K_d = 7.8(2)$ for $N_t=8$ 
%and 16).
The chiral limit, $1/K=4$, cannot be achieved in the confining phase 
even at $T=0$. 
Similar result for $N_F=7$ is given in Ref.~\cite{previo}.
Therefore, we conclude that the transition is a bulk transition.

From these results, we obtain the phase diagram at $\beta=0$
shown in Fig.~\ref{Kd.Nf}.
The first-order transition line separating the confining phase and the 
deconfining phase 
reaches $1/K=4$ (chiral limit for small $N_F$) between $N_F=6$ and 7.
Note that the critical value of the
inverse critical hopping parameter is greatly
increased from $1/K=4.08$ to 7.75 when $N_F$ increases from 7 to 240. 
%(This corresponds to the value of the 
%quark mass roughly from 0.25 to 1.0 in units of $a^{-1}$.)
Recall that $1/K=8$ is the point for massless free quarks at
$\beta = \infty$.

\begin{figure}[tb]
\begin{center}
a)\epsfile{file=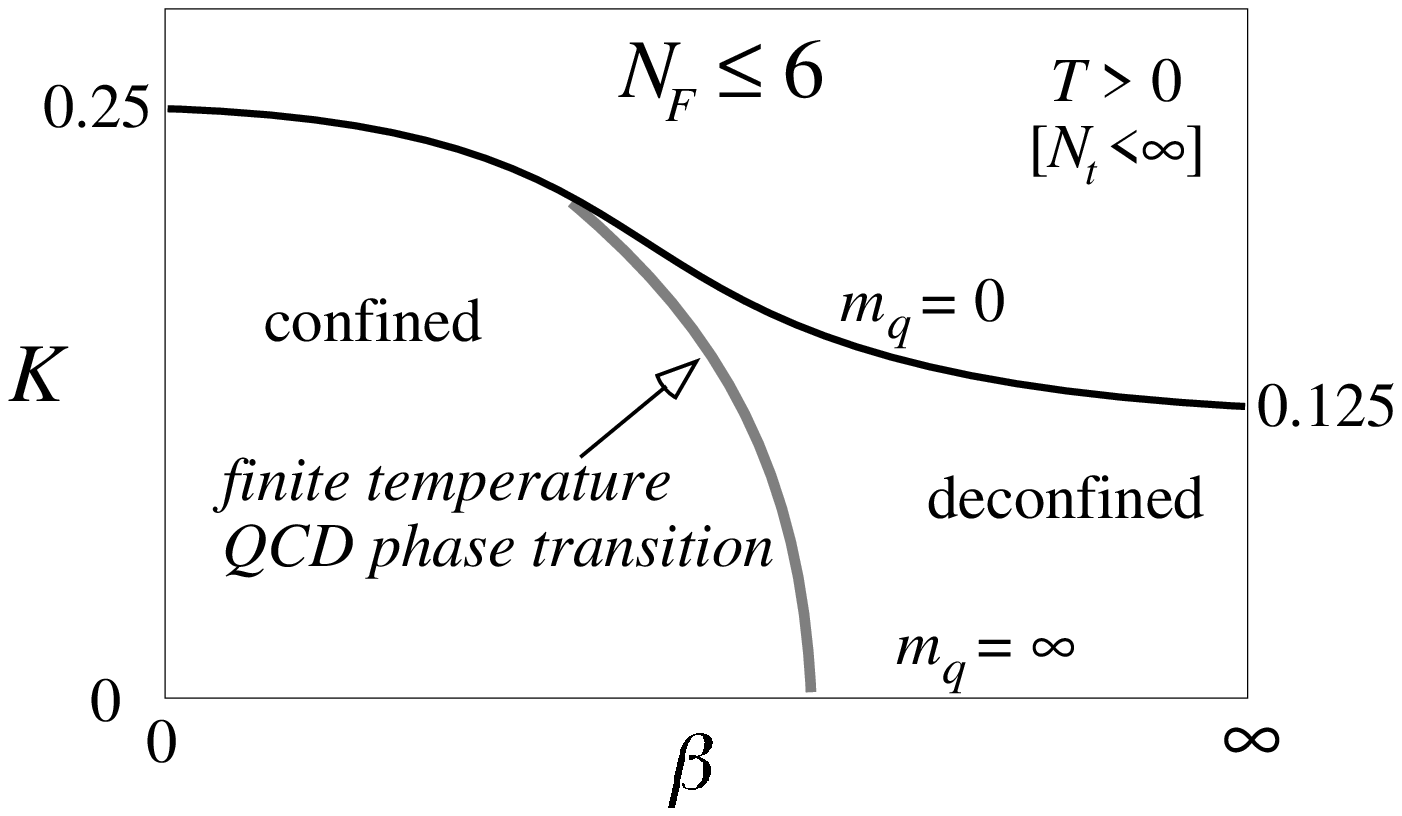,width=0.45\textwidth}
b)\epsfile{file=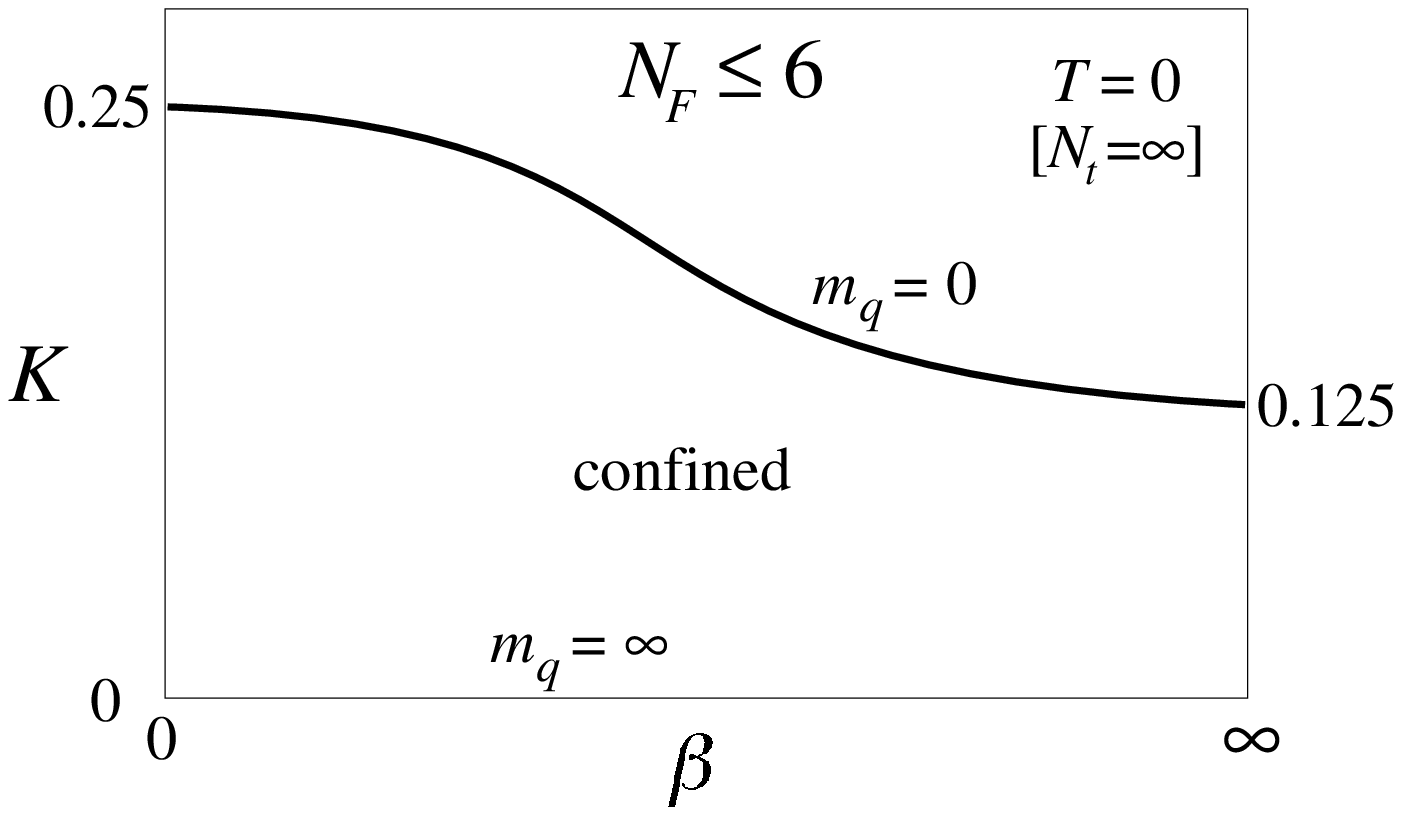,width=0.45\textwidth}
\end{center}
\vspace{-0.5cm}
\caption{The phase structure for $N_F \leq 6$
for (a) $T>0$, and (b) $T=0$.
}
\label{NfSmall}
\end{figure}

%%%%%%%%%%%%%%%%%%%%%%%%%%%%%%%%%%%%%%%%%%%%%%%%%%%%%%%%%%%%%%%%%%%%%
\section{Phase structure at finite $\beta$}
\label{sec:finitebeta}

Let us now study the case $\beta > 0$.
We perform simulations at $\beta = 0.0$--6.0
varying $N_F$ from 2 to 300.
On a lattice with a fixed finite $N_t$, we have the 
finite temperature deconfining transition 
when the quark mass is sufficiently heavy.
In particular, at $K = 0$, we have the first order deconfinement 
transition of the pure gauge theory at finite $\beta$
irrespective to the value of $N_F$.

%%%%%%%%%%%%%%%%%%%%%%%%%%%%%%%%%%%%%%%%%%%%%%
%\subsection{$N_F \leq 6$}

For $N_F \leq 6$, the finite temperature transition/crossover line 
crosses the $K_c$ line at finite $\beta$ \cite{Stand26}.
A schematic diagram of the phase structure for this case is shown 
in Fig.~\ref{NfSmall}. 
(In the followings, we call the transition/crossover line simply
as transition line.)
We have a monotonous RG flow towards the strong coupling
limit also on the $K_c$ line.
Therefore, the whole finite temperature transition line moves towards 
a larger $\beta$ as $N_t$ is increased.
In the limit $N_t=\infty$ (i.e., at $T=0$), only the confining phase 
is left.  

%%%%%%%%%%%%%%%%%%%%%%%%%%%%%%%%%%%%%%%%%%%%%%
%\subsection{$N_F \geq 7$}

\begin{figure}[tb]
\begin{center}
a)\epsfile{file=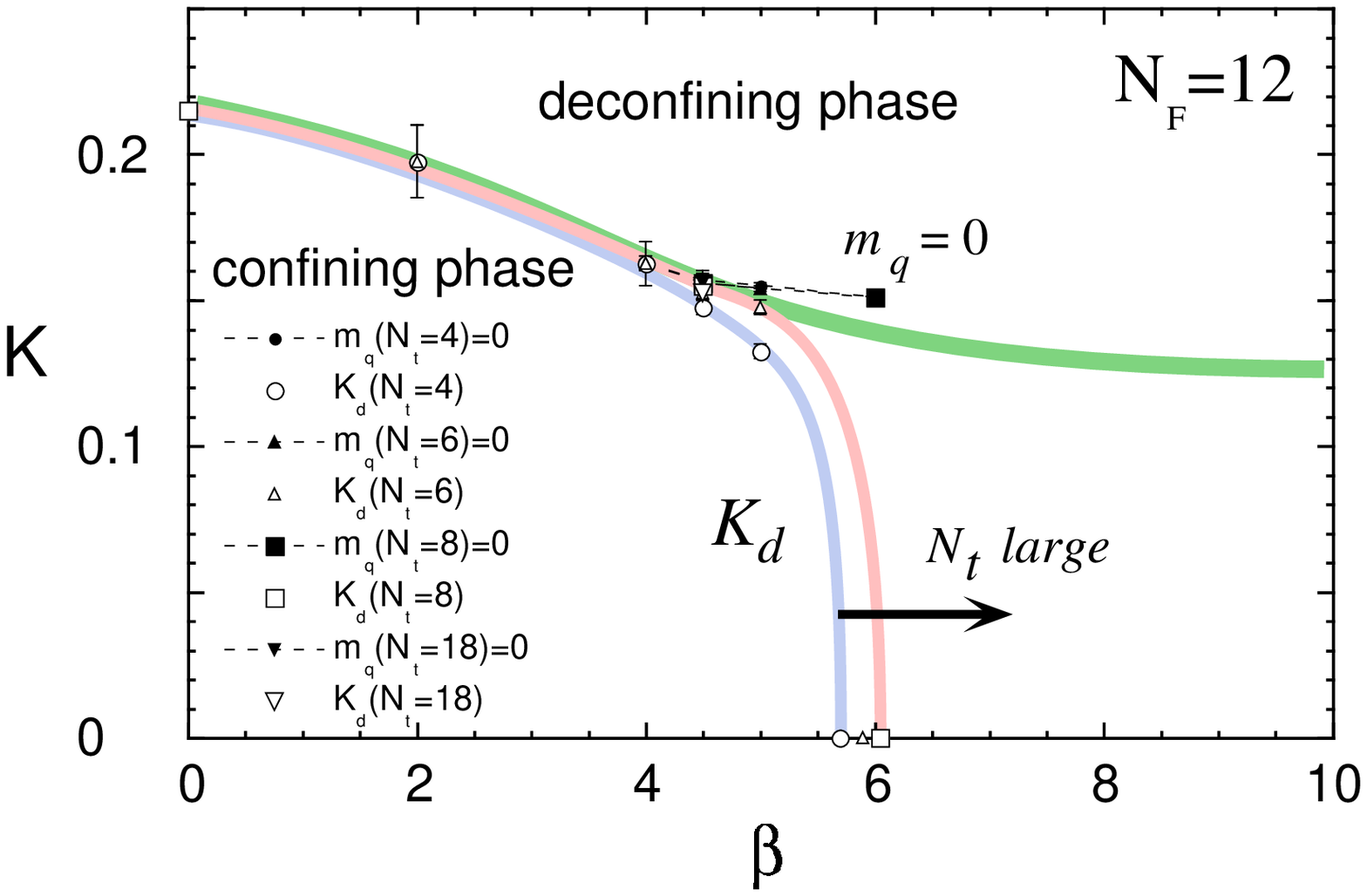,width=0.45\textwidth}
%\makebox[5mm]{}
b)\epsfile{file=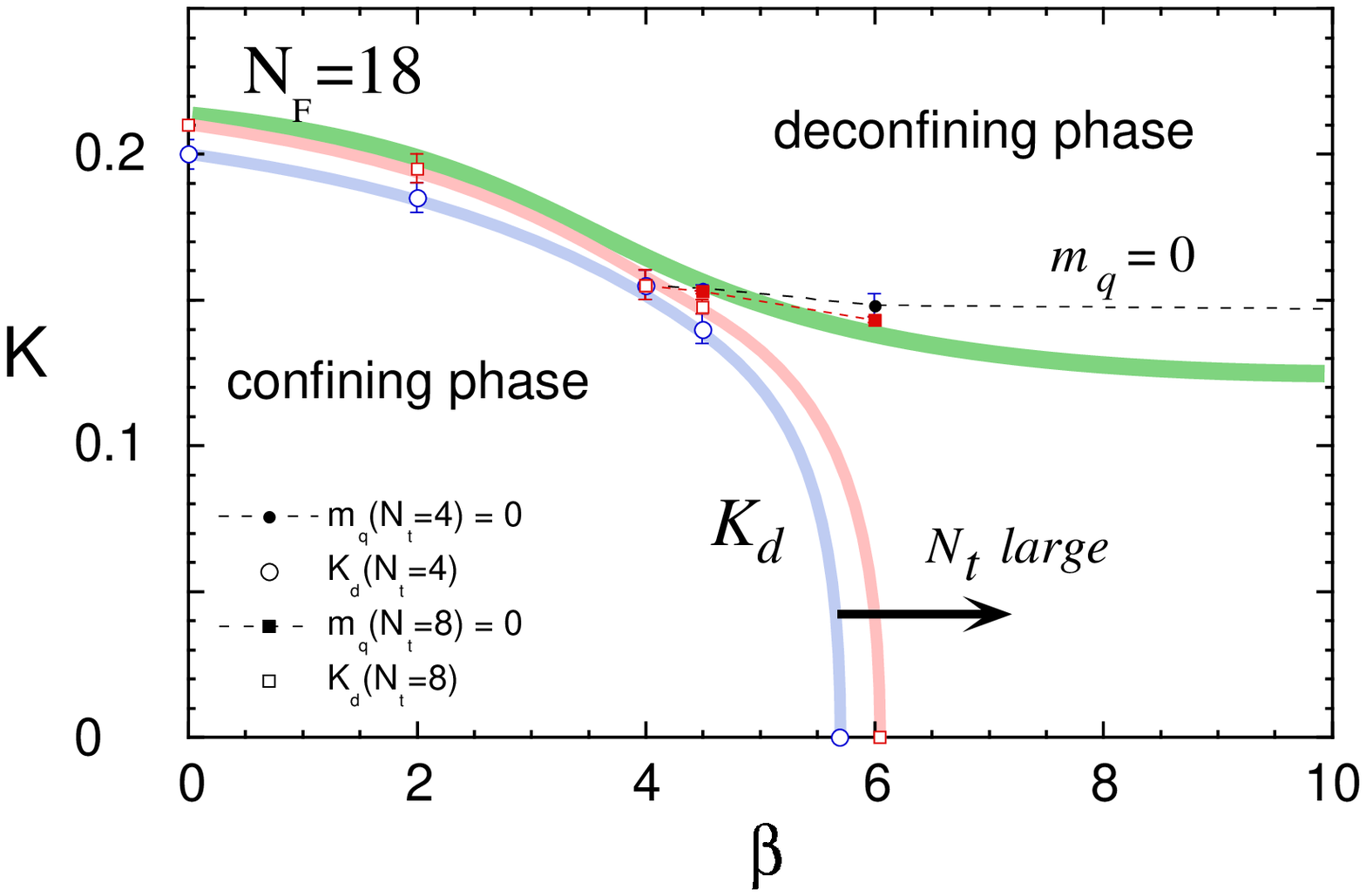,width=0.45\textwidth}
\end{center}
\vspace{-0.5cm}
\caption{
Phase diagram for (a) $N_F=12$ and (b) $N_F=18$.
Dark shaded lines represent our conjecture for the bulk transition 
line in the limit $N_t=\infty$.
}
\label{Nf12_18}
\end{figure}

\begin{figure}[tb]
\begin{center}
\epsfile{file=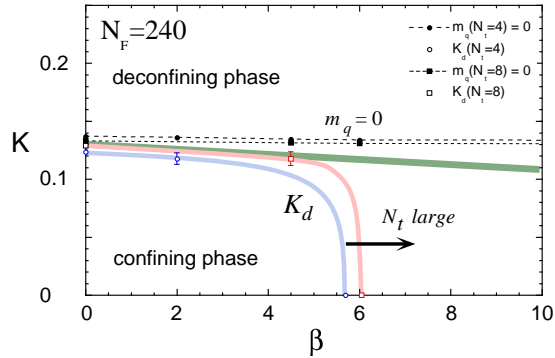,width=0.5\textwidth}
\end{center}
\vspace{-0.5cm}
\caption{Phase diagram for $N_F=240$.}
\label{Nf240}
\end{figure}

For $N_F \geq 7$, we have seen that there exist a bulk 
deconfining transition in the strong coupling limit.
We have to clarify the relation between
the finite temperature transition at small $K$ 
and the bulk transition at $\beta=0$.
From our simulations, we obtain the phase diagrams
sown in Fig.~\ref{Nf12_18} for $N_F=12$ and 18, and 
in Fig.~\ref{Nf240} for $N_F=240$.
When $N_t =4$ or 8, the phase boundary line between confining
and deconfining phase bends down at finite $\beta$
due to the finite temperature phase transition
of the confining phase.
The darker shaded lines are our conjecture for the phase boundary between 
the confining phase and the deconfining phase at zero temperature
($N_t=\infty$).

In Figs.~\ref{Nf12_18} and \ref{Nf240}, 
the dashed lines in the deconfining phase represents the 
chiral limit, $m_q=0$. 
This line also corresponds to the minimum point of $m_\pi^2$.
We have checked that, for the case $N_F=240$, the quark propagator 
in the Landau gauge actually shows the chiral symmetry, 
$\gamma_5 G(z) \gamma_5 = - G(z)$, on the $m_q=0$ line \cite{lat9496}. 
For $N_F \simm{>} 240$, we find that the $m_q=0$ line in the 
deconfining phase, which starts
at $1/K=8$ in the weak coupling limit $\beta=\infty$, 
reaches the strong coupling limit,
as shown in Fig.~\ref{Nf240} for $N_F=240$.
For a smaller $N_F \simm{<} 100$, 
because the bulk transition line $K_d$ in the strong coupling 
region shifts toward larger $K$ with decreasing $N_F$, 
the $m_q=0$ line in the deconfining phase
hits the $K_d$ line at finite $\beta$.
For example, in the case $N_F=18$, it hits at $\beta = 4.0$ -- 4.5,
as shown in Fig.~\ref{Nf12_18}(b).

%%%%%%%%%%%%%%%%%%%%%%%%%%%%%%%%%%%%%%%%%%%%%%
\subsection{RG-flow for $N_F \geq 17$}

The $m_q=0$ point at $\beta=\infty$ is a trivial IR fixed point
for $N_F \ge 17$. 
The phase diagram shown in Fig.~\ref{Nf240} suggests that 
there are no other fixed points on the $m_q=0$ line at finite $\beta$.

\begin{figure}[t]
\begin{center}
a)\epsfile{file=figure/nf240.mass.ps.2,width=0.5\textwidth}
%\makebox[5mm]{}
b)\epsfile{file=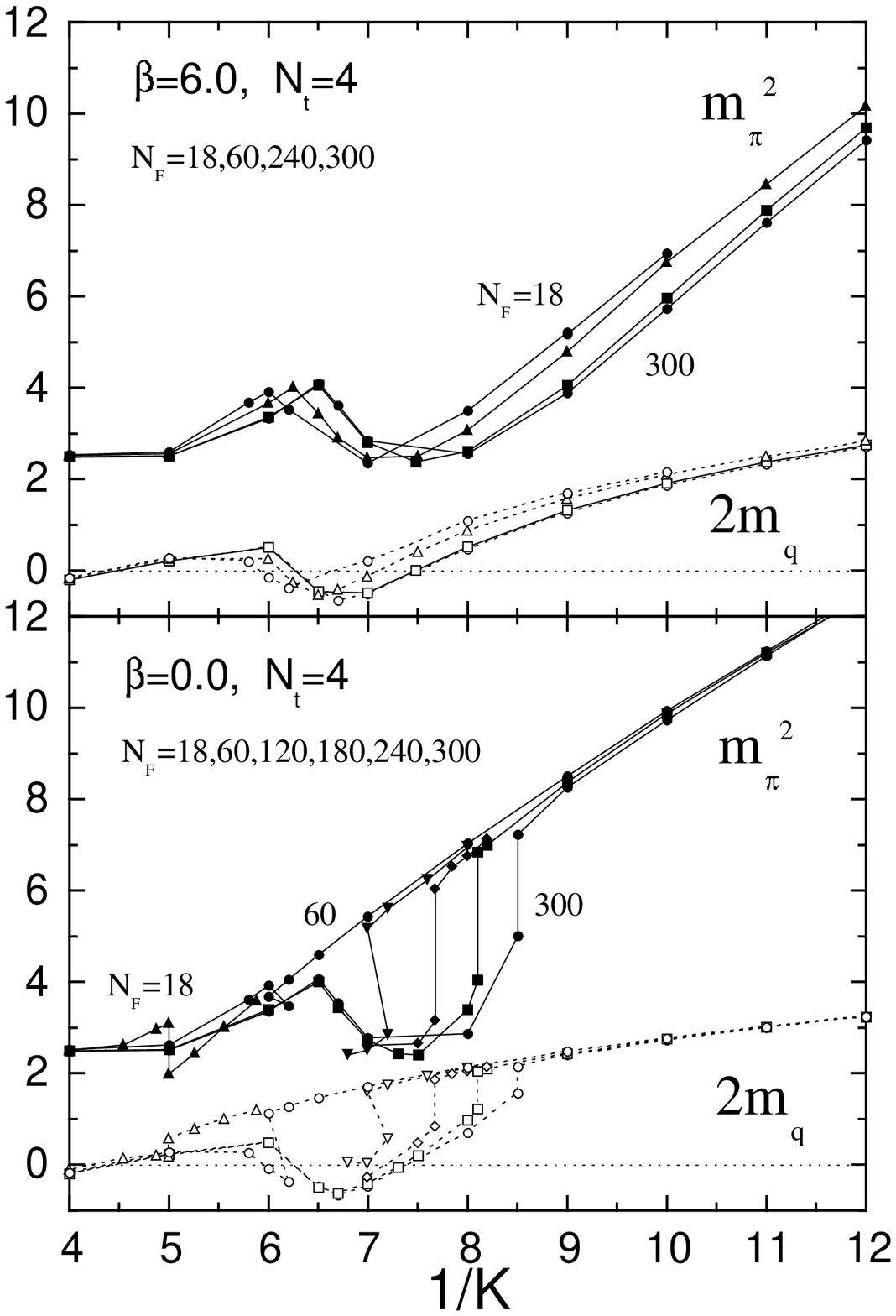,width=0.4\textwidth}
\end{center}
\vspace{-0.5cm}
\caption{
Results of $m_\pi^2$ and 2$m_q$ versus $1/K$ at $N_t=4$:
(a) $N_F=240$ at various $\beta$.
(b) $N_F=18$ -- 300 at $\beta=6.0$ and 0.0.
}
\label{massNfLarge}
%\vspace{-0.4cm}
\end{figure}

Figure~\ref{massNfLarge}(a)
shows the results of $m_\pi^2$ and $2m_q$ in the deconfining 
phase for $N_F=240$ at $N_t=4$.
We find that the shape of $m_\pi^2$ and $2m_q$ 
as a function of $1/K$
only slightly changes for $1/K < 8$
when the value of $\beta$ decreases from $\infty$
down 0. 
We obtain similar results also for $N_t=8$.
This suggests that quarks are almost free down to $\beta=0$ 
in the deconfining phase.
The results for $N_F=300$ are essentially the same, except for 
very small shifts of the transition point.

We make a Monte Carlo renormalization group (MCRG) study 
along the $m_q=0$ at $N_F=240$.
Performing a block transformation with scale factor 2,
we estimate the quantity $\Delta \beta = \beta(2a) - \beta(a)$:
We generate configurations
on an $8^4$ lattice on the $m_q=0$ points at $\beta=0$ and 6.0
and make twice blockings. 
We also generate configurations on a $4^4$ lattice 
and make once a blocking.
Then we calculate $\Delta\beta$ 
by matching the value of the plaquette at each step.
We obtain $\Delta \beta \simeq 6.5$ at $\beta=0$ and $10.5$ at $\beta=6.0$.%
\footnote{
In order to get a more precise value of $\Delta \beta$,
one has to make many steps of block transformations, together with 
a fine tuning of parameters of the block transformation,
and do matching using several types of Wilson loops.
We reserve elaboration of this point for future works.
For $N_F=240$, because the velocity of the RG flow is large, 
we will be able to obtain the correct sign and approximate value of
$\Delta\beta$ by a simple matching.
}
The value obtained from the two-loop perturbation theory is 
$\Delta \beta \simeq 8.8$ at $\beta = 6.0$.
The signs are the same and the magnitudes are comparable.
This suggests that the direction of the RG flow on the $m_q=0$ line
at $\beta=0$ and 6.0 is the same as that at $\beta=\infty$.
This further suggests that there are no fixed points at finite $\beta$.
These imply that the theory is trivial for $N_F=240$.

The area of the deconfining phase decreases with 
decreasing $N_F$ in the strong coupling region. 
However, the shape and the values of $m_\pi^2$ and $m_q$ are quite similar
when we vary $N_F$ from 300 down to 17.
Fig.~\ref{massNfLarge}(b) shows $m_\pi^2$ and $2m_q$ 
for $N_F=18$ -- 300 at $\beta=6.0$ and 0.
At $\beta=6.0$, the shapes of $m_\pi^2$ are almost identical to each 
other, except for a small shift toward smaller $1/K$ as $N_F$ is 
decreased.

These facts suggest that, for $N_F \geq 17$, the nature of physical 
quantities in the deconfining phase is almost identical to that 
observed for $N_F=240$ and 300, i.e., quarks are almost free.
Therefore, we conjecture that the direction of the RG flow at 
$\beta \simm{<} 6$ is identical to the case $N_F=240$.
Combining this with the perturbative result
that $\beta=\infty$ is an IR fixed point for $N_F \ge 17$, 
we conjecture that the RG flow along the massless quark line 
in the deconfining phase is uniformly directing to smaller $\beta$,
as in the case of $N_F=240$.
When this is the case, the theory has only one IR fixed point at 
$\beta=\infty$, i.e.\ the theory is trivial for $N_F \ge 17$.

%%%%%%%%%%%%%%%%%%%%%%%%%%%%%%%%%%%%%%%%%%%%%%%%%%%%%%%%%%%%%
\subsection{$16 \geq N_F \geq 7$}
\label{sec:nf167}

As discussed in Sec.~\ref{sec:strong},
quark confinement is lost for $N_F \ge 7$ at $\beta=0$.
We have performed simulations for $N_F=12$ and 7.
The phase diagram for $N_F=12$ is shown in Fig.~\ref{Nf12_18}(a).
At $\beta \simm{<} 6.0$ we studied, the gross feature of the phase diagrams 
for $N_F=7$ and 12 is quite similar to the case $N_F=18$ shown 
in Fig.~\ref{Nf12_18}(b). 
Physical quantities at $\beta \simm{<} 6.0$ also show behavior 
similar to that shown in Fig.~\ref{massNfLarge} for $N_F \geq 17$.
Therefore, we consider it probable that the direction of the RG flow 
in the deconfining phase at small $\beta$ is towards a larger $\beta$ 
as in the case of $N_F \geq 17$.
However, the direction of the RG flow at $\beta=\infty$ is opposite to that 
for $N_F \geq 17$ because the theory is asymptotically free for $N_F \le 16$.
This means that we have an IR fixed point somewhere at a finite value 
of $\beta$.
The continuum limit is governed by this IR fixed point. 

On the $m_q=0$ line around $\beta=4.5$,
we find that $m_\pi$ is roughly twice the lowest Matsubara frequency.
This implies that the quarks are not confined and almost free. 
Therefore, the anomalous dimensions are small at the IR fixed point, 
suggesting that the IR fixed point locates at finite $\beta$.
%(For an IR fixed point at $\beta=0$, 
%the anomalous dimension would be large or infinite.) 
%
Unfortunately, as shown in Fig.~\ref{Nf12_18}, 
the massless line hits the boundary in the coupling parameter space. 
Without much space in the strong coupling region where simulation is
easy, it is hard to study numerically the location of the IR fixed 
point. 
A detailed study of RG flow in a wider coupling parameter 
space is required.
We reserve these studies for future works.

%In summary, we conjecture an IR fixed point at finite $\beta$ 
%for $16 \geq N_F \geq 7$.
%The theory in the continuum limit is a non-trivial theory with 
%small anomalous dimensions, however, without confinement. 

\begin{figure}[tb]
\begin{center}
\epsfile{file=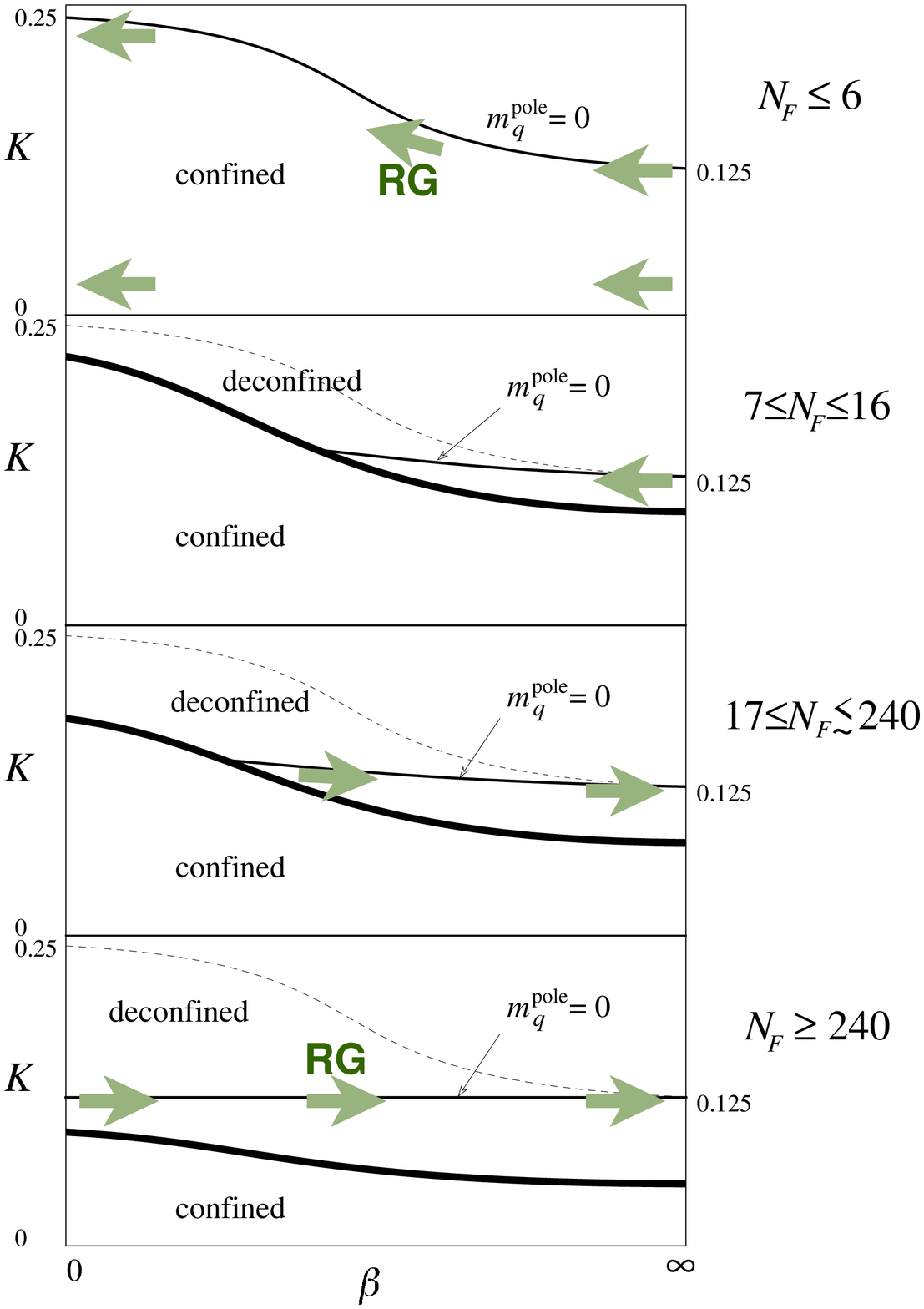,width=0.55\textwidth}
\end{center}
\vspace{-0.6cm}
\caption{Phase structure and the RG flow at $T=0$
for general number of flavors.
Thin dashed lines in the phase diagrams for $N_F \geq 7$ represent 
the location of the chiral limit for $N_F \leq 6$ as a guide for eyes.}
\label{RGflow}
\end{figure}

%%%%%%%%%%%%%%%%%%%%%%%%%%%%%%%%%%%%%%%%%%%%%%%%%%%%%%%%%%%%%%%%%%%
\section{Conclusions}
\label{sec:conclusions}

Summarizing our numerical results and conjectures, 
we propose Fig.~\ref{RGflow} for the phase structure of QCD at $T=0$:
When $N_F \le 6$, 
quarks are confined for any values of the current quark mass and $\beta$.
The chiral limit $K_c(\beta)$, where the current quark mass $m_q$ vanishes, 
belongs to the confining phase.
When $N_F \ge 7$, there is no chiral limit in the confining phase.
There is a line of a first order phase transition from the confining phase
to a deconfining phase at a finite current quark mass for all values
of $\beta$. 
The chiral limit exists only in the deconfining phase, 
and, therefore, can not be taken from the confining phase.

In order to see the orientation of the RG flow on the chiral limit line
in the deconfining phase, we performed a MCRG study at $N_F=240$, 
because, in this case, we have sufficiently long massless quark line 
in the strong coupling region where numerical simulation is easy. 
We found that the RG flow along the massless quark line at small $\beta$ 
is consistent with the result of the perturbation theory. This
suggests that the direction of the RG flow is uniform for 
all values of $\beta$, as shown in Fig.~\ref{RGflow}.
Therefore, there is only one IR fixed point at $\beta=\infty$, 
so that the theory in the continuum limit is trivial.
Accordingly, we find that, in the deconfining phase, 
the $K$-dependences of pion screening mass and current quark mass 
at small $\beta$ are almost identical to those of
free Wilson quarks.

Also for smaller values of $N_F \geq 7$, 
the general features of the phase diagram and the 
$K$-dependence of physical quantities 
are quite similar to the case of large $N_F \sim 240$.
Therefore, we conjecture that the direction of the RG flow is 
the same as in the case of $N_F=240$, 
at least for $\beta \simm{<} 6.0$ we have studied.

Together with the results of the perturbation theory, 
we conjecture the following picture.
\begin{itemize}
\item
For $N_F \geq 17$, because the RG flow in the weak coupling limit 
has the same direction due to the lack of asymptotic freedom, 
we conjecture a uniform RG flow towards the trivial IR fixed point,
as in the case of $N_F = 240$.
Thus the theory is trivial for $N_F \ge 17$.
\item
When $7 \le N_F \le 16$, because the theory is asymptotically free,
the direction of the RG flow is opposite at large $\beta$.
Therefore, we expect a non-trivial IR fixed point at finite $\beta$.
In this case, the theory in the continuum limit is a non-trivial 
theory with anomalous dimensions.
Because the coupling constant in the IR limit is finite, 
quarks are not confined in this case. 
\item
When $N_F \le 6$, 
quarks are confined and the chiral symmetry is spontaneously broken 
for any values of the current quark mass and $\beta$.
The chiral limit belongs to the confining phase.
\end{itemize}

\section*{ACKNOWLEDGEMENTS}

Numerical simulations are performed with 
HITAC S820/80 at KEK, and Fujitsu VPP500/30 and
QCDPAX at the University of Tsukuba.
This work is in part supported by the Grants-in-Aid of Ministry of 
Education, Science and Culture (Nos.~09304029 and 10640248).

%%%%%%%%%%%%%%%%%%%%%%%%%%%%%%%%%%%%%%%%%%%%%%%%%%%%%%%%%%%%%%%%%%%%%%%%%%%
%%%%%%%%%%%%%%%%%%%%%%%%%

\end{document}